\documentclass[aps,numerical,graphicx,reprint,pre]{revtex4-1}
\usepackage{amsmath}
\usepackage{graphicx}
\draft 

\begin{document}

%
%
%
\title{Fluctuation dissipation theorem and electrical noise revisited.}

\author{Lino Reggiani}
\email{lino.reggiani@unisalento.it}
\affiliation{Dipartimento di Matematica e Fisica, ``Ennio de Giorgi'',
Universit\`a del Salento, via Monteroni, I-73100 Lecce, Italy.}

\author{Eleonora Alfinito}
\affiliation{Dipartimento di Ingegneria dell' Innovazione, Universit\`a del
Salento, via Monteroni,  I-73100 Lecce, Italy}
\affiliation{Istituto Nazionale di Fisisca Nucleare (INFN), Sezione di Lecce, Italy.}
\date{\today}
%

\begin{abstract}
The fluctuation dissipation theorem (FDT) is the basis for a microscopic description of the interaction between electromagnetic radiation and matter.
By assuming the electromagnetic radiation in thermal equilibrium and the interaction in the linear response regime, the theorem interrelates the spontaneous fluctuations of microscopic variables with the kinetic coefficients that are responsible for energy dissipation.
In the quantum form provided by Callen and Welton in their pioneer paper of 1951 for the case of conductors (H.B. Callen and T.A. Welton, {\it Phys. Rev.} {\bf 83} (1951) 34-40), electrical noise detected at the terminals of a conductor, is given in terms of the spectral density of voltage fluctuations, $S_V(\omega)$,  was related to the real part of its impedance, $Re[Z(\omega)]$, by the simple relation:
$$
S_V(\omega)=2 \hbar \omega coth \left(\frac{\hbar \omega }{2K_BT}\right) 
Re \left[Z(\omega)\right]
$$
where $K_B$ is the Boltzmann constant, $T$ is the absolute temperature, $\hbar$ is the reduced Planck constant and $\omega$ is the angular frequency.
The drawbacks of this relation concern with: (i)  the appearance of a zero point contribution which implies a divergence of the spectrum at increasing frequencies; (ii)  the lack of detailing the appropriate equivalent-circuit of the impedance, (iii) the neglect of the Casimir effect associated with the quantum interaction between zero-point energy and boundaries of the considered physical system; (iv) the lack of identification of the microscopic noise sources beyond the temperature model.
These drawbacks do not allow to validate the relation with experiments. 
By revisiting the FDT within a brief historical survey since the genesis of its formulation, that we fix with the announcement of Stefan-Boltzmann law (1879-1884), we shed new light on the existing drawbacks by  providing further properties of the theorem with particular attention to problems related with the electrical noise of a two-terminals sample under equilibrium conditions. Accordingly, among the others, we will discuss the duality and reciprocity properties of the theorem, its applications to the ballistic transport regime, to the case of vacuum and to the case of a photon gas. 
\end{abstract}
\maketitle
\section{Introduction}
The fluctuation-dissipation theorem (FDT) is a pillar of statistical physics by interrelating the interaction between electromagnetic fields and matter.
In essence, it asserts that linear response of a given system to an external perturbation is expressed in terms of fluctuation properties of the system in thermal equilibrium. 
Even if the FDT is generally associated with the announcement of Nyquist relation in 1928 \cite{nyquist28},  its genesis can be traced back to the Stefan-Boltzmann law of 1878-1884 that relates the total power radiated by a black-body to the fourth power of the absolute temperature. 
Since then,  its formulation  received particular attention by many scientists. 
Yet, to date  several ambiguities and paradoxes  remain which strongly hide the relevant power of this theorem in interpreting a large variety of physical phenomena. 
The aim of this paper is to revisit the FDT,
and in particular its application to electrical noise, and shed new light on details and properties of the theorem that are usually not considered in the standard literature and/or are avoided because they inhibit to carry out a comparison between theory and experiments.
While we attempt to describe a wide range of phenomena, this selection is by no means exhaustive, and highly biased by subjective interests.
Significant examples are the individualization of the natural bandwidths of the fluctation spectra under classical conditions described by Nyquist relations, and the role of the zero-point contribution emerging under quantum conditions.
Together with several original contributions, the review collects many results already available in the literature, with the purpose to provide a unifying picture for the benefit of a deeper understanding of the interaction between light and matter at a microscopic and macroscopic level.  
We believe that this paper, by touching the most relevant aspects of the subject within an historical survey, can also serve as a reference for a broad audience of scholars and  researchers interested to further advance on the topic. 
\par
As physical system of interest we consider a cavity of length, $L$, and cross-sectional area, $A$, filled with a homogeneous medium with the terminal surfaces acting as ideal metallic contacts, and embedded in a thermal reservoir at a given temperature $T$.
When the medium is replaced by vacuum the standard black-body is recovered.
Here we mostly focus on: the instantaneous voltage fluctuations measured between the  open terminals of the physical system and the real part of its impedance or
the instantaneous current fluctuations measured in the short circuit connecting opposite  terminals of the system and the real part of its admittance.
The theoretical  formulation of the FDT in the time domain  implies an inter-relation between the correlation function, describing fluctuations of an observable (the noise), and the response function, describing the associated linear-response coefficient (the signal).
By Fourier transform, these functions can be expressed in the frequency domain (spectra) thus allowing for an easier comparison of theory with experiments.
Statistical mechanics under equilibrium conditions and linear-response theory to an external perturbation are the most rigorous way to provide a microscopic formulation of the FDT.
We want to stress, that among the other, the above formulation of the FDT provides a unified interpretation of the black-body radiation spectrum and of the thermal agitation of electric charge in conductors. 
\par
From an historical point of view, the following steps provide a time sequence of the most important achievements that can be traced back to the FDT \cite{haar54,parisi01,boya04}. 
\par\noindent
1 - (1860), Kirchhoff \cite{kirchhoff1860}, On the relation between the radiating and absorbing powers of different bodies for light and heat. \\
2 - (1879-1884), Stefan-Boltzmann law for the total emissivity of a black-body at a given temperature. \\
3 - (1896-1905),  Wien (1896), Rayleigh (1900), Planck (1901), Jeans (1905)
laws for the power spectrum of the black-body radiation. \\
4 -  (1905), Einstein \cite{einstein05} diffusion-mobility relation for a classical gas. \\
5 - (1908), Langevin \cite{langevin08} stochastic approach and fluctuation force. \\
6 - (1912), Planck \cite{planck12} inclusion of zero-point energy of a quantum oscillator. \\
7 - (1927), Johnson  \cite{johnson27} thermal  noise in conductors, experiments.  \\
8  - (1928), Nyquist  \cite {nyquist28} thermal  noise in conductors, theory. \\
9  - (1931), Onsager \cite{onsager31}  reciprocity relations of kinetic coefficients in irreversible processes. \\
10 -  (1948), Casimir \cite{casimir48} attraction between opposite metallic plates in vacuum. \\
11 - (1951), Callen-Welton \cite{callen51} quantum fluctuation dissipation theorem for conductors. \\
12 -  (1957), Kubo \cite{kubo66}  quantum formalism of correlation and response functions.\\
Recent advances in the field comprise: FDT in the  microcannical, canonical and grand-canonical ensembles, reciprocity and duality relations in the FDT, the role of Casimir effect in the black-body radiation spectrum 
\cite{bonanca08,reggiani16,reggiani17}. 
\par
According to the above historical survey, the first important manifestation of the FDT can be traced back from Stefan-Boltzmann black-body radiation law announced in the period  1879-1884, and relating light absorption and thermal radiation of a macroscopic body.
The theoretical interpretation of the black-body radiation spectrum involved the formulation of the Rayleigh-Jean law of 1900 and then of the celebrated Planck law of 1901. 
\par
A second manifestation of the FDT was  Einstein relation between mobility and diffusivity of 1905, that is at the basis of Brownian chaotic molecular-motion.
The theoretical interpretation of the Brownian motion led Langevin to develop in 1908 a new theoretical approach based on stochastic differential equations.
\par
A third manifestation of the FDT was the discovery by Johnson  of the spontaneous electrical fluctuations between the terminals of a conductor at equilibrium in 1927 and the subsequent theoretical formulation by Nyquist in 1928. 
\par
A fourth manifestation of the FDT was the prediction by Casimir in 1948 of the existence of an attractive force between two parallel metallic plates in vacuum, which was
associated  with the unavoidable presence of the zero-point fluctuations of the electromagnetic field.  This {\it quantum agitation}  is due to the uncertainty principle and represents the vacuum counterpart of thermal agitation.
Casimir prediction was later confirmed experimentally with increasing accuracy and theoretically justifiable within the quantum expression of the FDT formulated by Callen and Welton in 1951.
\par
Theoretical modeling of FDT can be traced back to the Rayleigh-Jeans law, followed by Wien and Planck laws for the interpretation of black-body radiation spectrum. Then, further developments include  Einstein relation and the Langevin formalism of stochastic differential equations, Nyquist relation for the interpretation of Johnson electrical noise, Callen and Welton quantum generalization of Nyquist relation, Kubo formalism and the use of a generalized quantum Langevin equation \cite{ford88}.
\par
The most rigorous theoretical derivation of the FDT makes use of quantum statistical mechanics.
To this purpose, fundamental advances started from Callen and Welton pioneer paper of 1951, where a first order perturbation theory was used, and Kubo operatorial formalism of 1957.
In both cases use was made of a canonical ensemble approach.
Bonanca 2008 \cite{bonanca08} derived the FDT for the case of a microcanonical ensemble and Reggiani et al \cite{reggiani16} for the case of a grand canonical ensemble, generalizing the results to the case of quantum fractional-statistics.
\section{Critical formulation of the fluctuation dissipation theorem}
The essence of the problem  addressed in this section can be formulated by considering the thermal noise-power per unit  bandwidth, $dP(f)/df= u(f,T)$, with $u(f,T)$ the energy spectrum at frequency $f$ and temperature $T$ radiated into a single mode of the electromagnetic field by a physical system coupled to a thermal reservoir.
As physical systems, here we consider the relevant cases of: (i) a one-dimensional harmonic oscillator and the fluctuations of the carrier displacement $X(t)$, (ii) a conductor and the fluctuations of the voltage drop at the open terminals $V(t)$, 
(iii) a conductor and the fluctuations of the current flowing in the short circuit $I(t)$.
We notice that for an ideal black-body cavity the impedance coincides with the vacuum resistance given by $R_{vac}=\mu_0 c$, with $\mu_0$ the vacuum permeability and $c$ the light speed in vacuum.
Thus, the black-body can be considered as a particular case of a resistor.
\par
For all the cases of interest $u(f,T)$ is a universal function  of frequency and temperature given by:
\begin{equation}
u(f,T) = S_X(f) \frac{2\pi f} {4 Im \{\alpha_x(f)\}} \,
\end{equation} 
\begin{equation}
u(f,T) = S_V(f) \frac{1}{4 Re\{Z(f)\}} \,
\end{equation} 
\begin{equation}
u(f,T) = S_I(f) \frac{1}{4 Re\{Y(f)\}} \,
\end{equation} 
where $S(f)$ is the (symmetrized) spectral-density of the fluctuations of the corresponding observable $X, V, I$, and  $\alpha_x(f)$ (the electrical susceptibility), $Z(f)$ (the impedance) and $Y(f)=1/Z(f)$ (the admittance) are  the response coefficients of the considered physical system.
\par
The spectral density and the associated response coefficient are the Fourier-Laplace  transform of the corresponding (symmetrized) correlation-function  and of the response function in the time domain.
Thus,  by interrelating the fluctuations with the dissipative part of the response, the above expressions for $u(f,T)$ represent the mathematical formulation of the fluctuation dissipation theorem. 
Both the correlation and the response functions can be expressed within a classical or a quantum formalism, as will be detailed in the following.
According to the literature, the thermal noise-power per unit bandwidth  takes three possible universal forms as:
\begin{equation}
u_{RJ,N}(f,T) =
K_BT  \ ,
\end{equation} 
\begin{equation}
u_{P}(f,T) =
K_BT \frac{x}{e^x-1} \ ,
\label{uplanck}
\end{equation} 
\begin{equation}
u_{P,CW,K}(f,T)=
K_BT  \frac{x}{2} coth(\frac{x}{2})=
u_{P}(f,T) + \frac{hf}{2} \ ,
\label{uzp}
\end{equation}
where $u_{RJ,N}(f,T)$ refers to Rayleigh-Jeans  and Nyquist \cite{nyquist28} classical formulation,  $u_{P}(f,T)$ refers to Planck 1901 \cite{planck01} quantum formulation  and $u_{P,CW,K}(f,T)$ refers to Planck 1912 \cite{planck12}, Callen-Welton 1951 \cite{callen51} and Kubo 1957 \cite{kubo66} quantum formulation including zero point contribution,  with
$x=hf/K_BT$ where $h$ is the Planck constant, and $K_B$ is the Boltzmann constant.
\begin{figure}
 \centering
  \includegraphics[width=\columnwidth]{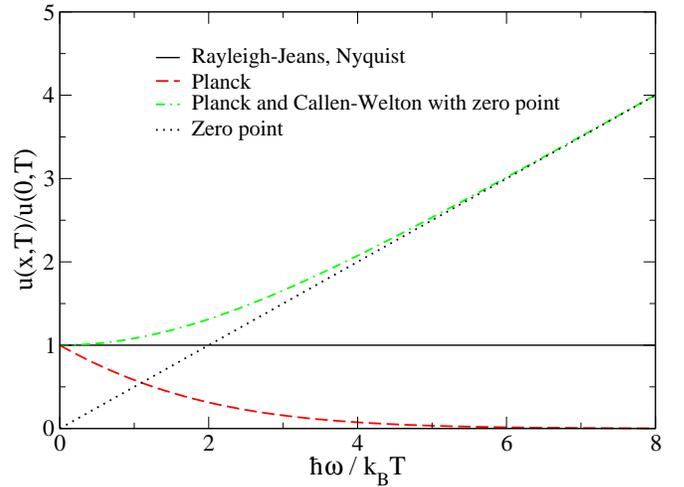}
 \caption{ 
Different shapes of $u(x,T)$ with $x=\hbar \omega/(K_BT)$ according to different models: black continuous curve refers to Rayleigh-Jeans and Nyquist, red dashed curve refers to Planck black-body, green dot-dashed curve refers to Planck, Callen and Welton, Kubo with zero-point contribution included, black dotted curve refers to zero-point only.}
\label{fig_spectra_u}
\end{figure}
We notice that Rayleigh-Jeans and Planck considered the black-body as the physical system of interest, while Nyquist, Callen and Welton, and Kubo considered basically a two terminal sample of given impedance (or admittance) without specifying the equivalent circuit, that is, the associated reactance.
All the three forms satisfy the conditio:
\begin{equation}
u(f=0,T) = K_BT \,
\end{equation}
a universal property valid for both classical and quantum physical systems.
\par
The total energy, $U(T)$, associated with a black-body of volume $V_0$ is defined as:
\begin{equation}
U(T) = \sum u(f,T) = V_0 \frac{8  \pi}{c^3} \int_0^{\infty}  f^2 u(f,T) df \,
\end{equation}
where the sum is extended over all the photon modes.
\par
We stress that only the Planck form, $u_P(f,T)$ leads to a finite value of $U(T)$,
in agreement with experiments and Stefan-Boltzmann law.
By contrast, both the other two forms lead to a so-called ultra-violet catastrophe.
We remark, that the split into two contributions of $u_{P,CW,K}(f,T)$, as given by the last expression in the r.h.s. of Eq. (\ref{uplanck}), is of most physical importance. Indeed,  after integration over all frequency range each contribution leads to the macroscopic and exclusive  value of the  total-energy associated with the spectrum.
\par
The first term is the Planck-contribution and represents a property of the coupling between the thermal reservoir and the physical system at thermal equilibrium. 
It follows from the detailed energy-balance related with the microscopic process of energy exchange between the thermal reservoir and the physical system. 
As such, it is a universal function of the temperature which takes a finite value at any frequency.
Accordingly, it vanishes at $T=0$,  it is independent of the external shape of the  physical system and of the conducting material inside to the physical system. Its spectrum can be directly measured by standard experimental techniques in a wide range of frequencies, typically from mHz to THz, and excellent agreement between theory and experiments is a standard achievement.
This contribution is the responsible of the thermal agitation at an atomic level of macroscopic bodies.
\par
By contrast, the second term, $hf/2$, represents a quantum property of the vacuum, and from its definition, it does not vanish at $T=0$. 
Its expectation value diverges (a so called vacuum catastrophe) but it is responsible of an attractive or repulsive force acting between opposite surfaces of a confined physical-system, as predicted by Casimir in 1948 \cite{casimir48} for the simple case of two thin parallel conducting plates in vacuum. 
This Casimir force is now thought to pertain to a more general family of so called fluctuation-induced forces that are ubiquitous in nature, covering many topics from biophysics to cosmology \cite{kardar99}. 
\par
The essential difference between the above two contributions is better explained when considering the total energy obtained by summation over all the photon modes.
For the Planck contribution, $U_{P}$, the summation is easily performed and gives the well-known result expressed by the Stefan-Boltzmann law for the total internal-energy associated with a black-body system at a given temperature:
\begin{equation}
U_{P}(T) = \sum K_BT \ \frac{x}{e^x-1} 
= \frac{2 \pi^5 V_0 (K_BT)^4}  { 15 c^3 h^3 } 
= 2.7 N_p K_B T \,
\label{usb}
\end{equation}
where the sum is extended over all the photon modes, and $N_p=2.02 \times 10^7 \ V_0 T^3$ is the expected number of photons inside the volume $V_0$ of the physical system at the given temperature.
We remark, that Eq. (\ref{usb}) expresses the Stefan-Boltzmann law in terms of 
thermal averages of the number of photons (modes) times the average photon energy 
in analogy with the case of a classical gas.
Here the difference is that the instantaneous number of photons is not conserved and their average number is a function of temperature.
Furthermore, being Bosons,  photons interact in terms of the symmetric properties of their wave-functions.
Accordingly, their bunching property is reflected in the coefficient $2.7$ slightly lower than the value of $3$ associated with the relativistic classical-massive case. 
\par
For the zero-point term, the same summation gives the expectation value of the zero-point total energy, $U_{zp}$, defined as 
\begin{equation}
U_{zp} = \frac{1}{2} \sum hf .
\label{eq:casimir}
\end{equation}
The sum gives a puzzling divergent energy contribution \cite{callen51,Dirac34} when evaluated all over the space.
Otherwise, as observed by Casimir \cite{casimir48}, real measurements are performed on finite-size systems where manifestations of zero-point energy are directly observable \cite{milton}.
In particular, the different content of electromagnetic energy inside and outside 
an assigned region produces a force that, in the case of two thin parallel conducting plates in vacuum, is attractive \cite{casimir48}.
In general, Eq. (\ref{eq:casimir}) is solved by using specific boundary conditions related to:  (i) the shape of the physical system, (ii) the material inside the physical system.
Calculations are not easy to be performed \cite{ederly06,schmidt08,auletta09}, and here we report the simple but significant  case considered by Casimir \cite{casimir48} and further confirmed by more detailed mathematical approaches \cite{milton}: 
\begin{equation}
U_{C} = - \frac{\pi A h c }{1440 L^3} .
\end{equation}
The negative value of the Casimir energy, $U_{C}$, corresponds to an attractive force (the Casimir force) between opposite conducting plates, $F_C$, given by
\begin{equation}
F_C
=- \frac{\pi h c A}  {480 L^4} .
\end{equation}
\par
As a consequence of this force, the physical system is not mechanically stable and the two opposite conducting plates forming the terminals would tend to implode \cite{lebowitz69} when left free to move.
Following standard mechanical arguments, to keep the stability a reaction vincular-force $F_{RV}= - F_C $, mostly ascribable to the rigidity of the physical system associated with its elastic properties \cite{kittel04}, should be introduced. 
We remark, that for macroscopic physical systems of centimeter length-scale as considered here, both forces take negligible values (of the order of $10^{-23} \ N$).
Accordingly, by accounting for the Casimir force and the associated vincular reaction, the resultant force is null, thus supporting the conjecture that at thermal equilibrium once mechanical stability is established zero-point energy cannot be extracted but its macroscopic effects are simply stored in the rigidity of the physical system.
\par
Indeed, the omission of the whole zero-point energy in considering black-body radiation spectrum is often encouraged for all  practical calculations
\cite{kogan96,prigogine14}. 
This omission can be justified by the fact that quantum agitation of vacuum does not interfere with particle thermal-agitation in a medium, rather  it can be exactly compensated by forcing the stability of the physical system.  
We shall therefore drop the zero-point contribution in the 
expression (\ref{uzp}) for the  energy spectrum  and recover the celebrated Planck distribution.
As a consequence, the original Planck 1901 \cite{planck01} expression for the black-body radiation emission as well as the Nyquist relation for the electrical noise in dissipative conductors which replaces $K_BT$  by the Planck distribution (as originally suggested by Nyquist himself \cite{nyquist28}) are justified, in full agreement with experimental evidence.
\section{Derivation  of fluctuation dissipation theorem within a microcanonical ensemble}
 Following \cite{bonanca08}, this scheme assumes an isolated system whose dynamics is given by the Hamiltonian $H$. 
An external harmonic perturbation of the form, $\hat V=- x f(t)$, is then applied to the system, where $f(t)$ is a classical force depending on time and $ x$ is the displacement operator.
\par
By taking the expression for the microcanonical density operator,
$\rho_m$, as:
\begin{equation}
\rho_m(E_t) =\frac{E_t -H}{Z(E_t)}
\end{equation}
where $Z(E_t)=Tr \delta(E_t-H)$, with $E_t$ the total energy of the system,
to derive the FDT use is made of  the  representation of $\delta(E_t - H)$ as inverse Laplace transform: 
\begin{equation}
\delta(E_t - H)=
\frac{1}{2 \pi i} \int_{\gamma - i \infty}^{\gamma + i \infty} exp[(E_t - H)z] dz\end{equation}
%
from which the quantum correlation function is introduced
\begin{equation}
C_{xx}(z,t)=
Tr(e^{-Hz}x(0)x(t))
\end{equation}
and the corresponding asymmetric spectral density
%
\begin{equation}
J_{xx}(\omega) =
\frac{1}{2 \pi}\int_{-\infty}^{\infty} C_{xx}(z,t) e^{-i\omega t}dt 
\end{equation}
satisfying the condition of detailed-energy balance: 
%
\begin{equation}
J_{xx}(z,-\omega)=exp(-\hbar\omega z)J_{xx}(z,\omega).
\end{equation}
By considering the macroscopic displacement due to the action of an harmonic potential, the symmetrized spectral-density of spontaneous fluctuations, $S_{xx}(z,\omega)$, and the Fourier transform of  the linear-response-function measured through the same observable $x$, $\chi_{xx}(z,\omega)$, are given by 
\begin{equation}
\biggl [{ {S_{xx}(z,\omega)} \atop { \frac{1}{i \hbar} \{\chi_{xx}(z,\omega)\} } }
\biggr ]
=
{1\over 2}
\biggl [ { {J_{xx}(z,\omega)+J_{xx}(z,-\omega)} 
\atop {J_{xx}(z,\omega) - J_{xx}(z,-\omega) } } \biggr ]
\end{equation}
From the above set of equations we obtain
\begin{equation}
S_{xx}(z,\omega) ={\frac{i \hbar}{2}}coth(\frac{z \hbar \omega}{2})  \ \chi_{xx}(z,\omega) \ 
\label{qfdt}
\end{equation}
which is the quantum FDT relating the spectra of the correlation function (i.e. fluctuations) and of the response function (i.e. dissipation). 
\par
In the classical limit $\hbar \rightarrow 0$ it is
\begin{equation}
S_{xx}(z,\omega) =\frac{i}{z \omega} \ \chi_{xx}(z,\omega) \
\label{cfdt}
\end{equation}
which is the classical FDT. 
\par
One easily realizes from Eqs. (\ref{qfdt}) and  (\ref{cfdt}) that the replacement of $z$
by $1/(K_BT)$ in those equations leads precisely to the quantum and classical versions of the FDT obtained in the canonical ensemble \cite{zubarev71,landau74}.
\par 
We remark that Eq. (\ref{qfdt}), by containing the zero-point contribution,  predicts the vacuum catastrophe already discussed in Sect (II).
As a consequence,  Eq. (\ref{qfdt}) is not appropriate to be compared with experiments.  Rather, ﬂuctuations in this case happen due to the dynamics of the concerned system itself and not due to the coupling to a thermostat as in the canonical ensemble.
In addition to the pure meaning of the relation between response and correlation, one may wonder whether Eqs. (\ref{qfdt}) and (\ref{cfdt}) can be useful or not. 
Although beyond the scope of the present work, a general and deep discussion of the subtle points mentioned above as well as of the linear response theory for the microcanonical ensemble, would remain of great interest and value.
\section{Classical Langevin approach}
Here we express the spectral density and the kinetic coefficients entering the definition of $u(f,T)$ within a classical stochastic approach introduced by Langevin \cite{langevin08}. 
To justify a one dimensional treatment and neglect magnetic effects we consider the condition $L^2 \ll A$, where an instantaneous number of carriers $N(t)$ are present in the physical system.
From total-current conservation and  Ramo-Shockley-Pellegrini theorem \cite{shockley38,ramo39,pellegrini86,pellegrini93} it is:
\begin{equation}
\frac{d}{dt} V(t) = \frac{L}{\varepsilon_0 \varepsilon_r A} \left[ \frac{q}{L} \sum_{i=1}^{N(t)} v_{i,x}(t) + I(t) \right]
\end {equation}
with $V(t)$ the instantaneous voltage drop at the sample terminals, $\epsilon_0$ and $\epsilon_r$ the vacuum permittivity and the relative dielectric constant of the medium, respectively, $q$ the unit charge, 
$v_{i,x}(t)$
the instantaneous longitudinal velocity of the $i$-th carrier inside the sample,  and $I(t)$ the instantaneous total-current.
\par
Then, a Drude-Langevin equation is added to describe the electron motions inside the system:
\begin{equation}
m \frac{d}{dt} v_d(t) + \frac{m}{\tau} v_d(t) = -\frac{q}{L} V(t) + \Gamma(t)
\end{equation}
where $v_d(t)$ is the instantaneous drift-velocity of carriers inside the sample, 
$m$ is a carrier effective mass, $\tau$ is the momentum scattering-time, and $\Gamma(t)$ is the instantaneous Langevin random-force satisfying,
\begin{eqnarray}
\overline{\Gamma(t)} &=& 0 \\
\overline{\Gamma(t) \Gamma(t^{\prime})} &=& \gamma \delta(t-t^{\prime})
\end{eqnarray}
where $\gamma$ is the strength of the noise source and  overline denotes ensemble average.
\par
By using the  plasma frequency $\omega_p$
\begin{equation}
\omega_p^2 = \frac{\overline{N} q^2}{\varepsilon_0 \varepsilon_r m A L}
\end{equation}
and the definition of the carrier drift-velocity 
\begin{equation}
v_d(t) = \frac{1}{N(t)} \sum_{i=1}^{N(t)} v_{i,x}(t) 
\end {equation}
where $v_{i,x}(t)$ is the instantaneous value of the $i$-th carrier velocity component along the $x$ direction, we obtain
\begin{equation}
\frac{d}{dt} V(t) = \frac{mL}{\overline{N}} \omega_p^2 \left[\frac{q}{L} N(t) 
v_{d}(t) +   I(t) \right] 
\end {equation}
Now, the lumped equivalent-circuit and the two extreme  boundary conditions associated with the detection of the electrical fluctuations are considered.
\subsection{Equivalent circuit and spectral densities}
\begin{figure}
\centering
\includegraphics [width=\columnwidth]{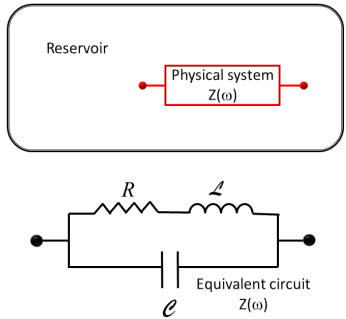}
 \caption{ Schematic of the statistical ensemble considered here.
The reservoir can be a grand-canonical or a canonical according to the 
operation conditions for noise detection at constant current or constant voltage 
operation modes, respectively.
The equivalent circuit of the intrinsic impedance $Z(\omega)$ consists of a resistor $R$ with length L 
and cross sectional are A, a  kinetic-inductance ${\cal L}$   and a parallel 
plates capacitor $\cal {C}$ filled with the homogenous medium that constitutes the resistor of given relative dielectric-constant.
The capacitance  and inductance  account
for the presence of the contacts and for the inertia of carriers, respectively.
See text for details.
}\label{figec}
\end{figure}

The equivalent circuit of a real resistor (conductor) which is consistent with the Langevin approach is shown in Fig. (\ref{figec}).
Here the Ohmic resistance independent of frequency is coupled with the intrinsic capacitance $\cal {C}$, that accounts for contact effects, and the intrinsic kinetic-inductance ${\cal L}$,  that accounts for inertial effects of charge carriers.
Accordingly, the impedance (admittance) $Z(\omega)$ ($Y(\omega)$) of the equivalent circuit is given by
\begin{equation}
Z(\omega) = \frac{1}{Y(\omega)} =
R \frac{1+i \omega \tau}{1 - \omega^2 \tau_p^2+i \omega \tau_d}
\label{eq1}
\end{equation}
\begin{equation}
Re[Z(\omega)] = 
R \frac{1}{(1 - \omega^2 \tau_p^2)^2+ \omega^2 \tau_d^2}
\label{eq1}
\end{equation}
\begin{equation}
Re[Y(\omega)] =
G \frac{1}{1 + \omega^2 \tau^2}
\label{eq1}
\end{equation}
where: the resistance is given by
\begin{equation}
R = \frac{1}{G} = \frac{L^2 m}{q^2 \overline{N} \tau}
\end{equation}
with $G$ the conductance.   The kinetic inductance is given by
\begin{equation}
{\cal L} =\frac{A}{L}\mu_0 \mu_k = \frac{L^2 m}{q^2 \overline{N}}
\end{equation}
with $\mu_k$ the relative kinetic-permeability, and the capacitance is given by
\begin{equation}
{\cal C} = \frac{A}{L}\epsilon_0 \epsilon_r
\end{equation}
As usual,  the dielectric relaxation-time $\tau_d$, the scattering time $\tau$ and the plasma time $\tau_p$ are respectively given by:
\begin{equation}
\tau_d={\cal C}R, \ \tau={\cal L}/R, \ \tau_p^2={\cal CL} \,
\end{equation}
Notice that for centimetric length-scale of the physical system and for a standard carrier density of $10^{22} \ cm^{-3}$ for a conductor material, the order of magnitude of $\cal C$ and $\cal L$ are respectively of $10^{-13} \ F$ and $10^{-19} \ H$. Therefore, in experiments parasitic capacitance and inductance should play a significant role, unless high resistivity materials are considered.
\par
We remark that  $Re[Y(\omega)]$ decays as a Lorentzian with the characteristic momentum relaxation-time.
By contrast, $Re[Z(\omega)]$ decays 
with a spike at the plasma frequency, usually determining a plasmonic noise \cite{reggiani13}, which is washed out if the condition $\tau_d > \tau_p$ is satisfied. 
These spectra  were validated by Monte Carlo simulations \cite{reggiani13},
even if, because of the ultra-high frequencies involved (typically over the THz range) a direct experimental confirmation is still lacking. 
\par
The above equivalent circuit is consistent with the standard energy-equipartition relations
\begin{equation}
{\overline{V^2}}{{\cal C}} = 
{\overline{I^2}}{{\cal L}} 
= K_BT ,
\label{variances}
\end{equation}
which in this form are valid for any type of capacitance and inductance in the
circuit reported by Fig. (\ref{figec}). As such, this equivalent circuit is of most physical importance and should replace alternative equivalent circuits (like,
e.g., simple $R \cal C$ parallel and $R{\cal LC}$ series circuits) which are
also sometimes used in the literature but that do not provide plausible 
physical-spectra.
\subsection{Current noise operation}
In the current operation mode it is $V(t)=0$,  carriers enter and exit from the contacts and the current noise is detected in the external short-circuit.  The physical system should be considered an open one and is statistically described within a grand-canonical ensemble (GCE).
Accordingly,  the total current reads
\begin{equation}
I(t)= - \frac{q}{L} N(t)v_d(t)
\end{equation}
and the Drude-Langevin equation writes
\begin{equation}
\frac{d}{dt}I(t) + \frac{1}{\tau} I(t) = \frac{\overline{N}q}{Lm} \Gamma(t)
\end{equation}
\par
In full agreement with the equivalent circuit shown in Fig. (\ref{figec}), the Drude-Langevin equation can also be written as
\begin{equation}
\frac{d}{dt}I(t) + \frac{R }{\cal L} I(t) = 
\frac{L}{q \cal L} \Gamma(t) 
\end{equation}
\par
By using standard techniques \cite{mcquarrie}, from the homogeneous equation the current correlation-function writes
\begin{equation}
C_I(t) 
= \overline{I(0)I(t)} 
= \overline{I^2} e^{-t/\tau}
\end{equation}
and, from the inhomogeneous equation and energy equipartition, it is
\begin{equation}
\overline{I^2}= \frac{\overline{N}^2 q^2} {2 m^2 L^2} \gamma \tau
=\frac{K_BT}{\cal L}
\label{I2}
\end{equation}
Accordingly, the FDT relates $\gamma$ and $\tau$ as
\begin{equation}
\gamma \tau = \frac{2m K_B T}{\overline{N}}
\end{equation}
\par
From statistics, in the GCE the longitudinal diffusion-coefficient, generalized to include the effective interaction among carriers due to the symmetry properties of their wave functions and thus for the correct statistics \cite{gurevich79}, is given by
%
\begin{equation}
D
=\overline{v^{2'}_x} \tau
=\frac{K_B T \tau}{m} \frac {\overline {N}}  {\overline {\delta N^2}} 
\label{difcoef}
\end{equation}
where $\overline{v^{2'}_x}$ is the differential (with respect to carrier number) quadratic velocity component along the $x$-direction given by
\begin{eqnarray}
\overline{v^{2'}_x} 
= \sum_{\bf{k}} v^{2}_x  \,
\frac{\partial f(\varepsilon_{\bf{k}})} {\partial \mu_0}
\frac {\partial \mu_0} {\partial \overline{ N}} =
\frac{K_B T}{m} \frac {\overline {N}}  {\overline {\delta N^2}} \ .
\label{vx2p}
\end{eqnarray}
with $\mu_0$ the chemical potential.
\par
The main conclusion is that current fluctuations are associated with  fluctuations of the total  number of carriers inside the system that satisfy the Langevin equation:
\begin{equation}
\frac{d}{dt} \delta N(t) + \frac{1 }{\tau} \delta N(t) = 
\frac{L}{\sqrt{\overline{v^{2'}_x}}}\Gamma(t) ,
\end{equation} 
with the corresponding correlation function
\begin{equation}
C_{\delta N} (t) 
= \overline{\delta N(0) \delta N(t)} 
= \overline{\delta N^2} e^{-t/\tau}
\end{equation}
and from statistics in the GCE it is
\begin{equation}
\overline{\delta N^2}
=K_BT \frac{\partial \overline N}{\partial \mu_0}
\label{N2}
\end{equation}
\par
From Wiener-Khinchin theorem, the current spectral-density reads
\begin{equation}
S_I(\omega) = \frac{4 \overline{I^2}}{{\Delta f_I} } \ \frac{1}{1+(\omega \tau)^2} 
= 4 K_B T Re\{Y(\omega)\}
\label{SI}
\end{equation}
Equation (\ref{SI}) identifies $\Delta f_I = 1/\tau$ as the natural bandwidth of the current fluctuations spectrum and together with Eqs. (\ref{I2}) and (\ref{difcoef}) recovers a microscopic definitions of conductance:
\begin{equation}
G 
= \left(\frac{q}{L}\right)^2 D  \frac{\partial \overline{N}} { \partial \mu_0}
= \frac{q^2 \overline{v^{2'}_x} \tau} {L^2K_BT} \  \overline{ \delta N^2}
\label{GDR}
\end{equation}
which gives the generalized Einstein relation and expresses the conductance (i.e. dissipation) in terms of the variance of total carrier-number fluctuations (i.e. fluctuation), thus representing a microscopic form of the FDT  within a GCE.
\subsection{Voltage noise operation}
In the voltage operation mode it is $I(t)=0$, carriers remain inside to the physical system and the voltage noise is detected between the terminals of the external open circuit.  The physical system should be considered a closed one  and it is statistically  described   within a canonical ensemble CE, i.e. $\overline{N}=N=\textrm{constant}$.
Accordingly,  conservation of total current becomes
\begin{equation}
\frac{d}{dt}V(t)= \frac{mL}{q} \omega_p^2 v_d(t)
\end{equation}
and the Drude-Langevin equation reads
\begin{equation}
\frac{d^2}{dt^2}V(t) + \frac{1}{\tau} \frac{d}{dt} V(t) + \omega_p^2 V(t)= \frac{L}{q} \omega_p^2 \Gamma(t)
\end{equation}
In full agreement with the equivalent circuit shown in Fig. \ref{figec}, the Drude-Langevin equation can also be written as
\begin{equation}
\frac{d^2}{dt^2}V(t) + \frac{R}{\cal L} \frac{d}{dt} V(t)+ 
\frac{1}{{\cal LC}} V(t) = \frac{L}{q{\cal LC}} \Gamma(t)
\end{equation}
By using standard techniques \cite{mcquarrie}, from the homogenous equation  we obtain the correlation function
$$
C_V(t) = \overline{V(0)V(t)} 
$$
\begin{equation}
= \overline{V^2} \left[
\frac{ e^{-\lambda_1 t} \ \lambda_2}{\lambda_2 - \lambda_1} + 
\frac{ e^{-\lambda_2 t} \ \lambda_1}{\lambda_1 - \lambda_2} \right]
\end{equation}
with
\begin{equation}
\lambda_{1,2}= \frac{1}{2\tau} \pm \sqrt{\frac{1}{4\tau^2} -\omega_p^2}
\end{equation}
For $2 \omega_p \tau < 1$ the two real values of $\lambda$ describe a dumped  
behavior of $C_V(t)$, for   $2 \omega_p \tau > 1$  the two complex conjugate values of $\lambda$ describe an oscillating behavior of $C_V(t)$.
\par
From the inhomogeneous equation and energy equipartition, it is
\begin{equation}
\overline{V^2}= \frac{L^2 \omega_p^2}{2 q^2}  \gamma \tau
= \frac{K_BT}{\cal C}
\end{equation}
Accordingly, the fluctuation-dissipation theorem relates $\gamma$ and $\tau$ as
\begin{equation}
\gamma \tau = \frac{2m K_B T}{\overline{N}}
\end{equation}
\par
From statistics, in the CE it is
\begin{equation}
\overline{\delta v_d^2} =
\frac{1}{N^2} \sum_{\bf{k}} v_x^2 \  \overline{\delta f^2(\varepsilon_{\bf{k}})} 
= \frac{K_B T}  { m \ N} \ ,
\label{Deltavddef}
\end{equation}
where $\overline{\delta v_d^2}$ is the variance of the fluctuations of the
instantaneous carrier drift-velocity inside the system, $v_x=\hbar k_x
/ m$ is the $x$-component of its velocity, $\bf {k}$ is the carrier wave
vector, $\varepsilon_{\bf{k}}$ the corresponding energy and $\overline{
\delta f^2}$ is the variance of the equilibrium distribution-function $f$
normalized to carrier number which, according to statistics, satisfies the
property
\begin{equation}
\overline{\delta f^2(\varepsilon_{\bf{k}})} = -K_B T \frac{\partial f(\varepsilon_{\bf{k}})}
{\partial \varepsilon_{\bf{k}}} \ ,
\end{equation}
and, using the symmetry of the problem, $v_x^2$ is replaced by
$(2\varepsilon_{\bf{k}}/md)$, where $d$ denotes the dimension of the system.
With the density-of-states of a $d$-dimensional carrier gas satisfying ${\cal
D}(\varepsilon) \propto \varepsilon^{(d/2)-1}$, Eq.~(\ref{Deltavddef}) is
independent of dimensionality.
\par
Accordingly, we find
\begin{equation}
\overline{V^2} 
= \left(\frac{L m \omega_p} {q}\right)^2  \ \overline{\delta v_d^2} 
\label{V2}
\end{equation}
\par
From the equivalent circuit of Fig. (\ref{figec}), the spectral density reads
$$
S_V(\omega)
= \frac{4 \overline{V^2}}{\Delta f_V}  \ \frac{1}{(1- \omega^2 \tau_p)^2 +  (\omega \tau_d)^2}
$$
\begin{equation}
= 4 K_B T  Re\{Z(\omega)\}
\label{SV}
\end{equation}
Equation (\ref{SV}) interpolates between the oscillating and damped behaviors of the correlation function of voltage fluctuations and identifies $\Delta f_V = 1/ \tau_d$ as the natural bandwidth of the voltage fluctuations spectrum.
From  Eqs.  (\ref{Deltavddef}) and  (\ref{V2}) the following microscopic definition of resistance is obtained:
\begin{equation}
R = \frac{L^2 m^2} {q^2 \tau K_BT } \overline{\delta v_d^2} .
\label{Rvd2}
\end{equation}
which gives  the resistance (i.e. dissipation) in terms of the variance of carrier drift-velocity fluctuations (i.e. fluctuation), thus representing a microscopic form of the FDT  within a CE.
\par
We conclude, that voltage fluctuations are associated with fluctuations of carrier drift-velocity inside the system. 
\subsection{Damped harmonic oscillator}
The Langevin equation for the damped harmonic oscillator is analogous to that of  voltage fluctuations where voltage is substituted by carrier displacement along $x$ direction, i.e. $V(t) \rightarrow x(t)$, and the plasma frequency by the natural oscillator frequency, i.e. $\omega_p \rightarrow \omega_0$. 
The analogy is completed by noticing that the equipartition  law writes $ k \overline{x^2} =  K_B T$, $k$ being the strength of the oscillator (spring constant) and $\overline{x^2}$ being the  average squared-displacement along $x$ direction. 
Accordingly, the Langevin equation for the damped harmonic oscillator with $\omega_0 =\sqrt{k/m}$ writes:
\begin{equation}
{d^2 \over dt^2} x  + \gamma {d \over dt} x + \omega_0^2 x  = \frac{1}{m}\Gamma(t)
\label{1}
\end{equation}
\par
The corresponding spectral density is 
\begin{equation}
S_{xx}(\omega) = \frac{4  K_BT}{\omega}  Im\{\alpha_x(\omega)\}
\label{s_x}
\end{equation}
with $\alpha_x(\omega)$ the electrical susceptibility 
\begin{equation}
\alpha_x(\omega) = 
\frac{m }{\omega_0^2 - \omega^2 +  i (\gamma \omega)}
\label{3}
\end{equation}
\begin{figure}
 \begin{center}
  \includegraphics[width=9cm]{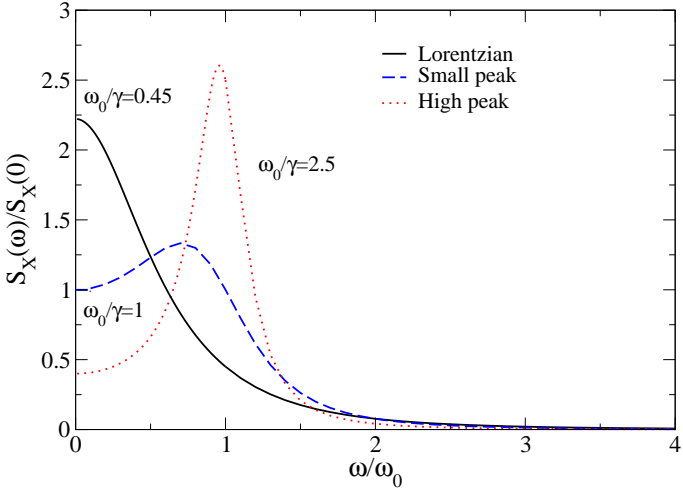}
 \end{center}
 \caption{ Shape of the normalized spectra of the damped harmonic oscillator for the cases:
$\omega_0 / \gamma=0.45$ continuous line, $\omega_0  / \gamma =1$ dashed line, and $\omega_0 / \gamma=2.5$ dotted line.
}
\label{spectra_dho}
\end{figure}
Figure (\ref{spectra_dho}) reports the shape of the spectra for the harmonic oscillator in the three cases: 
$\omega_0 / \gamma=0.45$, $\omega_0  / \gamma =1$, and $\omega_0 / \gamma=2.5$.
The spectra give evidence of the shift  from a Lorentzian shape when $\omega_0 < \gamma$ to a narrow peak centered at the natural oscillator-frequency when $\omega_0 > \gamma$ as predicted by Eq. (\ref{s_x}). 
\subsection{Ballistic regime}
The ballistic regime is controlled by the condition that the carrier  transit time between opposite contacts, $\tau_L$,  becomes shorter than the scattering time $\tau$.
Then,  by assuming for simplicity a one dimensional geometry,  at equilibrium the single-carrier longitudinal  velocity, $v_{x,b}$ inside the system remains constant in modulus. 
\par 
In an open system, $v_{x,b}$ is controlled by the injection velocity-distribution.
Following Landauer \cite{landauer99}, conductance becomes synonymous of transmission, i.e. all  carriers that are injected at one contact  are transmitted to the outside of the opposite contact.
The value of the ballistic conductance, $G^b$, becomes independent of the length of the system and is given by
\begin{equation}
G^b= \frac{1}{R^b}
=\frac{A q^2 n }{m \overline{v_{x,b}} }
\label{GRb}
\end{equation}
with $R^b$ the ballistic resistance and $n$ the carrier concentration.
\par
The ballistic resistance relies on a closed system where the ballistic carrier with a given $v_{x,b}$ is spectacularly reflected at the contacts. 
The transit time keeps the same value as for the case of conductance, but now resistance becomes synonymous of elastic reflection at the contacts and its value satisfies the reciprocity relation implied by Eq. (\ref{GRb}).
\par 
For the case of a classical one-dimensional injection distribution, it is 
\begin{equation}
\overline{v^c_{x,b}} 
= \sqrt{\frac{2 \pi K_BT}{ m}}
= \frac{L}{\tau^c_{L}}
\end{equation}
with $\tau^c_{L} =L/ \overline{v^c_{x,b}}$ the  transit time due to classical conditions, and for the equivalent circuit of Fig. \ref{figec}  it is
\begin{equation}
\tau^c_d = R^{b,c}C= \frac{L^2}{{\lambda_D}^2} \tau^c_{L}
\end{equation}
\begin{equation}
\tau^c = \frac{\cal L}{R^{b,c} }= \tau^c_{L}
\end{equation}
\begin{equation}
\tau^c_p = \frac{L}{\lambda_D }\tau^c_{L}
\end{equation}
with 
\begin{equation}
\lambda_D = \sqrt{\frac{\epsilon_0 \epsilon_r K_BT}{ n q^2} }
\end{equation}
the Debye length.
\par
For the case of  a degenerate one-dimensional injection distribution, it is \cite{greiner00}
\begin{equation}
\overline{v^d_{x,b}}  =   v_F 
= \sqrt\frac{2 \epsilon_F}{m}
= \frac{L}{\tau^d_{L}} 
\end{equation}
with $v_F$ and $\epsilon_F=\hbar^2 n_{1D}^2/(2m)$ the one dimensional Fermi velocity and Fermi energy, respectively, $n_{1D}$ the one dimensional carrier concentration, and $\tau^d_{L} =L/v_F$ the  transit time associated with degenerate conditions.
\par
For the equivalent circuit of Fig. \ref{figec}  it is
\begin{equation}
R^{b,d} = \frac{1}{G^{b,d}} = \frac{h}{e^2}
\end{equation}
per unit spin, and
\begin{equation}
\tau^d_d = R^{b,d}{\cal C} = \frac{L^3}{4 \alpha cA}
\end{equation}
\begin{equation}
\tau^d = \frac{{\cal L}}{ R^{b,d}} =  \frac{4 \alpha L^3}{c A}
\end{equation}
\begin{equation}
\tau^d_p =  \frac{ L^6}{c^2 A^2}
\end{equation}
with $\alpha=1/137$ the fine structure constant and
$L^3/(cA)$ a geometrical-averaged relativistic transit-time associated with  degenerate conditions.
\par
We remark, that under ballistic conditions, since the absence of scattering makes the  dynamics deterministic,  the correlation functions of current fluctuations  do not obey a simple  exponential relaxation but a more complicated decay depending from the carrier statistics.
According to \cite{greiner00}, the one-dimensional correlation functions of current fluctuations take the form, respectively for the ballistic-classical 
regime and the ballistic-degenerate regime:
\begin{equation}
\frac{C_I^{b,c}(x)}{C_I^{b,c}(0)} = erf(x^{-1}) - x[1-exp(-x^{-2})] ,
\end{equation}
and 
\begin{eqnarray}
\frac{C_I^{b,d}(x)}{C_I^{b,d}(0)} &=& 1-x \ for \ x \le 1 \\
\nonumber
&=& 0 \ for \ x>1 ,
\end{eqnarray}
with $x$ the time normalized to the respective ballistic transit-time.
\par
\begin{figure}
 \centering
\includegraphics[width=\columnwidth]{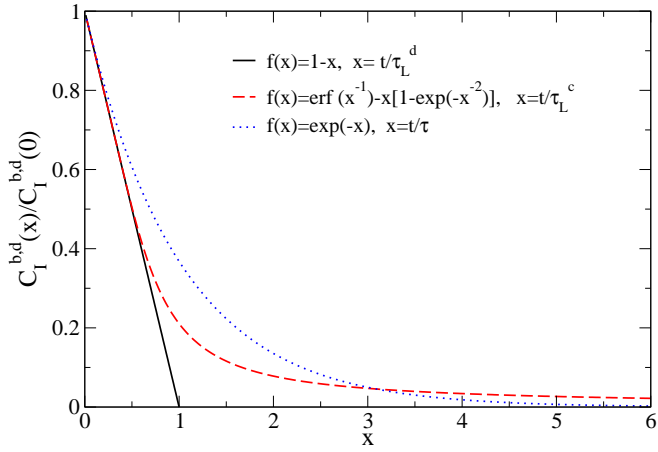}
 \caption{ Shape of the correlation functions of current fluctuations normalized to its initial value at $t=0$ in the case of one-dimensional ballistic regime.
Continuous curve refers to degenerate ballistic-conditions, dashed curve  refers to classical ballistic-conditions,  dotted curve refers to the case of an exponential decay typical of a diffusive regime.
}
\label{cf_ballistic}
\end{figure}

Figure (\ref{cf_ballistic}) reports the one-dimensional correlation functions of current fluctuations  given above.
Here curve 1 refers to the case of classical  ballistic-conditions and the correlation function is found to exhibit a long-time limit decay as $t^{-3}$.
Curve 2 refers to the case of degenerate ballistic-conditions, and the decay is found to exhibit a universal triangular shape. Curve 3 refers to the case of an exponential decay typical of a diffusive regime and is reported  for the sake of comparison.  
\par
The longitudinal diffusion coefficient takes a form analogous to that of the diffusive regime, but with the differential quadratic-velocity replaced by the quadratic injection-velocity and the scattering time replaced by the ballistic transit-time.
Thus, under ballistic conditions  diffusion like conductance becomes a global quantity that depends on the sample length as
\begin{equation}
D^{c,d}_b = \overline{v^{c,d}_{x,b}}^2 \tau^{c,d}_L  = \overline{v^{c,d}_{x,b}} L
\label{eq1}
\end{equation}
\subsection{Classical electrodynamics in vacuum}
By considering the case of vacuum, for the classical electric-field averaged over the black-body length, $E=V/L$, the corresponding Langevin equation is obtained from that of voltage fluctuations and reads:
\begin{equation}
\frac{d^2}{dt^2}E(t) + \frac{1}{\tau_E} \frac{d}{dt}E(t) + \frac{1}{\tau_E^2}E(t) = \Gamma_E (t)
\end{equation}
Using the vacuum definitions of resistance, capacitance and inductance  there is a single time scale, $\tau_E=L/c$ and the  correlation function takes the analytical form
$$
C_E(t) 
= \overline{E(0)E(t)}
$$
 \begin{equation}
= \overline{E^2} e^{-t/2 \tau_E}\times
[a sin( \frac{\sqrt3}{2} \frac{t}{\tau_E}) + 
b cos( \frac{\sqrt3}{2} \frac{t}{\tau_E})] 
\label{ctE}
\end{equation}
with $a= 1/4(\sqrt{3 \pi /2} +\sqrt{\pi /6})= 0.75$ and 
$b= \sqrt{\pi/2}=1.25$. 
\par
From classical energy-equipartition and FDT it is
\begin{equation}
 \overline{E^2} 
= \frac{K_BT  }{LA \epsilon_0}
= \frac{K_BT  R_{vac}} {A \tau_E}
= \frac{\gamma_E \tau_E }{2}
\label{E2}
\end{equation}
\par
For the classical magnetic-field averaged over the black-body geometry, 
$H= (L/A) I$ the corresponding Langevin equation is obtained from that of current fluctuations and reads:
\begin{equation}
\frac{d}{dt}H(t) + \frac{1 }{\tau_H} H(t) = 
\Gamma_H(t) 
\label{SH}
\end{equation}
with $\tau_H=\tau_E= L/c$ the correlation time of magnetic-field fluctuations.
\par
The  correlation function reads
\begin{equation}
C_H(t) = \overline{H(0)H(t)} 
= \overline{H^2} e^{-t/ \tau_H}
\label{ctH}
\end{equation}
\par
From classical energy-equipartition and FDT it is
\begin{equation}
 \overline{H^2} 
= \frac{K_BT} {LA \mu_0}
= \frac{K_BT G_{vac}} {A \tau_H}
= \frac{\gamma_H \tau_H }{2}
\end{equation}
\par
\begin{figure}
 \centering
\includegraphics[width=\columnwidth]{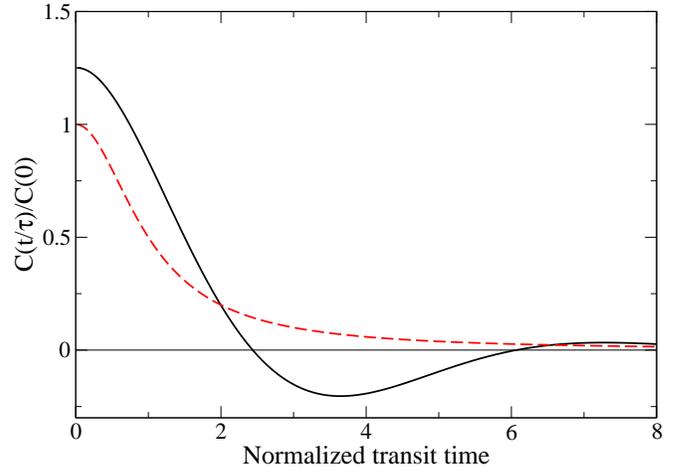}
 \caption{ Correlation functions of the  fluctuations of the classical electromagnetic fields in vacuum normalized to their initial value at $t=0$. Continuous curve refers to the electric field and dashed curve to the magnetic field, respectively.
}
\label{cf_vacuum}
\end{figure}

Figure (\ref {cf_vacuum}) shows the correlation functions of the electric-field and magnetic-field fluctuations in vacuum  given in Eqs. (\ref{ctE}) and (\ref{ctH}), respectively.
\par
For the spectral density of the electric-field fluctuations it is:
\begin{equation}
S_{E}(\omega) 
=  \frac{4 K_B T  R_{vac}}{L^2} \frac{1}{1-(\omega \tau_E)^2 + (\omega \tau_E)^4}
\label{SE}
\end{equation}
whose shape is the same of that of the damped harmonic oscillator reported in Fig. (\ref{spectra_dho}) when $\omega_0 = 1/\gamma = 1/\tau_E$.
\par
For the spectral density of the magnetic-field fluctuations it is:
\begin{equation}
S_{H}(\omega) 
=  \frac {4 K_B T  G_{vac}L^2}{A^2} \frac{1}{1+(\omega \tau_H)^2}
\end{equation} 
whose spectrum is a Lorentzian with $G_{vac}=1/R_{vac}$.
\par
We notice that, under voltage operation mode, only the fluctuations of the electric field are detected, while for current operation mode only the fluctuations of the  magnetic field  are detected.
\section{Duality and reciprocity of fluctuation dissipation theorem}
The dual property of electrical transport in the linear-response regime
asserts that perturbation (applied voltage  or imposed current)  and
response (measured current or voltage-drop) can be interchanged with the
associated kinetic coefficients (resistance or conductance,
respectively), being reciprocally interrelated. 
According to Ohm law, for a homogeneous conductor the dual property gives
\begin{equation}
V = R I \quad \mbox{and} \quad I = GV  ,
\label{IGV}
\end{equation}
that implies the reciprocity relation 
\begin{equation}
RG= 1 \ .
\label{GR}
\end{equation}
An analogous dual property and reciprocity relation can be formulated for electrical fluctuations at thermal equilibrium. 
Here the perturbation is the  source of the microscopic  fluctuations inside
the physical system (carrier number or carrier drift-velocity), and the  response is associated with the variance of the macroscopic fluctuating quantity
(current or voltage).
The individuation of the noise sources at a kinetic level,
and thus beyond the simple thermal-agitation model, is a major issue in statistical
physics that received only partial, and sometimes controversial, answers even
in the basic literature
\cite{johnson27,nyquist28,callen51,kubo66,klimontovich87}. 
\par
For the analysis at a kinetic level of current or voltage fluctuations  a
correct  definition of the physical system becomes of primary importance. 
On the one hand, the microscopic interpretation of carrier transport implies  the definition of an appropriate equivalent circuit at the macroscopic level. 
On the other hand, the detection of current or voltage fluctuations in the outside circuit should be related to the boundary conditions associated with the choice of the operation mode of detection (i.e. constant current or constant voltage) and the corresponding statistical ensemble of reference. 
Accordingly, current noise is measured in the outside short-circuit,
which implies an open system that is permeable to carrier exchange with the thermal reservoir, thus referring to a grand canonical ensemble (GCE).
By contrast, voltage noise is measured in the outside open-circuit which implies that carrier number in the system remains rigorously constant in time,
thus referring to a canonical ensemble (CE). While it is well-known that in
the thermodynamic limit different statistical ensembles become equivalent
\cite{landsberg54}, this does not hold anymore in the case of fluctuations,
where a finite-size system has to be considered. Nevertheless, we will show
that the dual property provides a simple relation between the noise sources acting  in the GCE and those acting in the CE.
Typical examples of physical system of interest  have been reported in the previous sections, where  use was made of a standard Langevin approach.
Here, the duality and reciprocity  properties of the FDT are addressed and
formally solved in the framework of the basic laws of statistical mechanics.
\par
According to the reciprocity property of the linear-response coefficients and their definitions in terms of the microscopic noise sources given in Eqs. (\ref{GDR})  and (\ref{Rvd2}) it is:
\begin{equation}
GR = \frac{\overline{v^{2'}_x} \overline{\delta N^2} m^2 \overline{\delta v_d^2}   } {(K_BT)^2  }
= \frac{\overline { v^{2'}_x}} {\overline{\delta v_d^2}} \
\frac{\overline{\delta N^2}}{N^2} = 1
\label{GR2}
\end{equation}
Thus, the microscopic noise sources satisfy the duality relation
\begin{equation}
\overline{ \delta N^2}  \ \overline { v^{2'}_x}  =
{ N}^2  \ \overline{ \delta v_d^2} \
= \frac{NK_BT}{m}   .
\label{N2v2}
\end{equation}
\par
For the variance of current  fluctuations, substitution of Eq.~(\ref{vx2p})
into Eq.~(\ref{I2}) gives the equivalent expressions:
\begin{equation}
\overline{ I^2}
= \frac{ q^2  \delta\overline{ N^2}}{\tau_N^2}
= \frac{A^2}{L^2}\overline{ H^2} ,
\label{eq14}
\end{equation}
with $\tau_N= \sqrt{L^2 / \overline{v^{2'}_x}}$
an effective transport-time through the sample \cite{greiner00} determining the conversion of carriers total-number fluctuations inside the sample into total-current  fluctuations measured in the external short-circuit, and with $\overline{H^2}$ the variance of magnetic-field fluctuations at equilibrium averaged over the geometry  of the physical system.
From Eq. (\ref{eq14}), the expression 
\begin{equation}
\overline{ H^2}
= \frac{L^2}{A^2}\overline{I^2}  ,
\label{eq14bis}
\end{equation}
represents a generalized Biot-Savart law that converts current  fluctuations inside the sample into fluctuations of the magnetic-field in proximity of the lateral surface of the sample averaged over the system  geometry.
\par
For the variance of voltage fluctuations, substitution of Eq.~(\ref{Deltavddef})
into Eq.~(\ref{V2}) gives the equivalent expressions
\begin{equation}
\overline{V^2} 
=\frac{L^2 \overline{\delta v_d^2}} {\mu_p^2} 
= L^2 \overline{E^2}
\label{eq16}
\end{equation}
with $\overline{ E^2}$ the variance of electric-field fluctuations at equilibrium averaged over the sample length,
and $\mu_p$ the plasma  carrier-mobility  given by
\begin{equation}
\mu_p=\frac{q \tau_p}{m}.
\label{eq17}
\end{equation}
Equation (\ref{eq16}) represents a generalized Ohm law that converts carrier 
drift-velocity  fluctuations inside the sample into electric-field  (or voltage) fluctuations at the terminals of the open circuit,
in analogy with the relation given in Eq. (\ref{eq14}) for the conversion of carrier-number fluctuations into magnetic-field fluctuations.
\par
Equations (\ref{GR2})  and (\ref{N2v2}) express the reciprocity and duality
properties of the microscopic noise-sources associated with the
fluctuation-dissipation relations in a medium. In other words, at thermodynamic
equilibrium carrier total-number fluctuations inside a conductor under
constant-voltage conditions are inter-related to carrier  drift-velocity
fluctuations under constant-current conditions.
Equations (\ref{eq14bis})  and (\ref{eq16}) express the reciprocity and duality
properties of the microscopic noise-sources associated with the
fluctuation-dissipation relations in the electromagnetic-field representation. 
\par
The dual property of the macroscopic FDTs is obtained from
Eqs.~(\ref{SI}), (\ref{V2}) and (\ref{N2v2}) as
\begin{equation}
\overline{I^2} = G^2_p \ \overline{V^2} \quad \mbox{and} \quad
\overline{V^2} = R^2_p \ \overline{I^2} ,
\label{I2V2}
\end{equation}
with the plasma conductance $G_p=(qN\mu_p)/L^2$ and the plasma resistance $R_p$
satisfying the reciprocity relation
\begin{equation}
G_p R_p=1 \ .
\label{GpRp}
\end{equation}
By satisfying the relations (\ref{SI}) and (\ref{SV}), the expressions
(\ref{I2V2}) and (\ref{GpRp}) justify the identification of the natural
bandwidths of the noise spectral densities assumed  here.
\par
For the variances of the  electromagnetic fields the duality relation writes:
\begin{equation}
\frac{\overline{E^2}}{\overline{H^2}} 
=\frac{A^2 \epsilon_r} {L^2  \mu_k} c^2.
\label{E2H2}
\end{equation}
\par
Equations (\ref{GpRp}) and (\ref{E2H2}) express the duality and reciprocity
properties of fluctuation-dissipation relations between the variances of the electric and magnetic fields associated with the correspondent kinetic coefficients.
All the above expressions hold for any type of statistics (in the case of
Bosons for temperatures above the critical temperature for Bose-Einstein
condensation \cite{davies68}), thus complementing the standard FDT in the
limit of low frequencies. From statistics, the two boundary conditions are
associated with a GCE and a CE, respectively, and Eq. (\ref{N2v2}) shows the
interesting results that both statistics provide the same result even outside
the thermodynamic-limit conditions \cite{landsberg54}.
\section{Fractional exclusion statistics}
The fractional exclusion statistics \cite{halperin84,wu95} generalizes the quasi particle distribution function to the case in which the statistical factor, $g$, continuously spans the range of values $0 \le g \le 1$ so that \cite{wu95}
\begin{equation}
f(x,g) = \frac{1}{W(x,g) +g}
\label{eq-fes}
\end{equation}
where $x=(\epsilon_k . \mu_0)/K_BT$, and $W(x,g)$ satisfies the implicit equation
\begin{equation}
W^g  (1+W)^{1-g} = exp(x)
\label{eq-W}
\end{equation}
The limiting values $g=0$ and $1$ correspond to Bose-Einstein and Fermi-Dirat statistics, respectively.
The variance of the fractional exclusion distribution follows the general definition of the GCE, and is given by:
\begin{equation}
\overline{\delta f(g,x)^2} = K_BT \left(\frac{\partial f} 
{\partial \mu_0}\right)_T  = -\frac{\partial f}{\partial x}
\label{eq-fes}
\end{equation}
which formally is the same as those previously used for the Bose-Einstein and Fermi distributions. 
\par
We conclude that present results are valid also in the case of fractional statistics.
\section{Black-body radiation spectrum}
The Langevin random-term contains the temperature as source of the external random-force and the friction coefficient (relaxation rate) as source of dissipation. From quantum electrodynamics, the temperature is basically related to the black-body radiation spectrum at thermal equilibrium, with the angular frequency, $\omega =2 \pi f$, being related to the photon energy $\epsilon_p=\hbar \omega$ and the classical equipartition law for the energy spetrum radiated into a single mode being substituted by Planck law as
\begin{equation}
K_BT \rightarrow K_BT \frac{x}{exp(x) -1}
\end{equation}
with $x=\hbar \omega / K_BT$.
\par
The above spectrum corresponds to replace the delta function of the time correlator of the Langevin random-force with a correlation function obtained by Fourier inverse transformation of the Planck factor as \cite{gardiner00}
\begin{equation}
\delta(t-t') \rightarrow \frac{1}{\tau_{\epsilon}}  (\frac{1}{y^2} - cosech^2(y)) 
\label{cf_planck}
\end{equation}
with $y=(t-t') / \tau_{\epsilon}$ and $\tau_{\epsilon} = \hbar / (K_BT)$ the 
quantum thermal (energy) correlation-time.
(Notice that at $T=1 \ K$ it is $\tau_{\epsilon} = 7.2 \ ps$, thus comparable with a scattering time.)
\begin{figure}
 \centering
\includegraphics[width=\columnwidth]{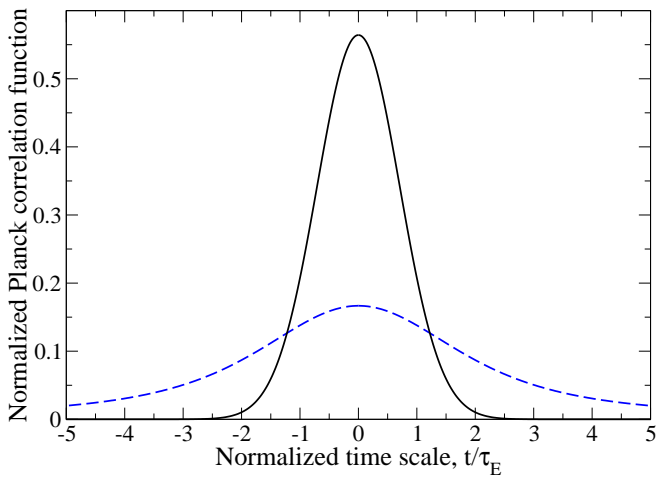}
 \caption{ Correlation function of the Langevin random-force.
Continuous curve refers to the normalized Planck black-body law of relation (\ref {cf_planck}) and dashed curve refers to the normalized Gaussian shape $(1/\sqrt{\pi}) exp(t/\tau_{\epsilon})$. Both curves look significantly broadened with respect to the classical delta-function shape, with the Planck curve exhibiting a longer tail than the Gaussian curve .Correlation function of the Langevin random-force.
}
\label{planck_cf}
\end{figure}
\par
Figure (\ref{planck_cf}) shows the normalized correlation function associated with the Planck law (continuous curve) which is compared with a normalized Gaussian function (dashed cure).  
Both curves look significantly broadened with respect to the shape of the delta function corresponding to the classical condition, with the Planck correlation-function exhibiting a longer tail than that of the Gaussian function. 
\par
\begin{figure}
 \centering
\includegraphics[width=\columnwidth]{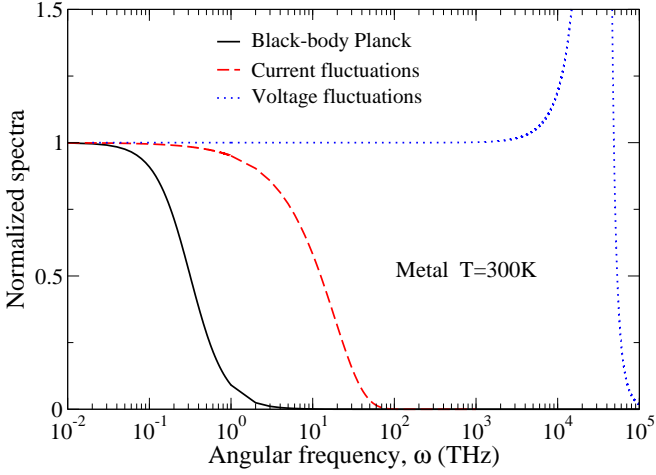}
 \caption{Noise and black-body spectra near cut-off for the case of a typical metal at $T=300 \ K$. Here typical time scales are, respectively, of $\tau=10^{-2} \ ps$, $\tau_d=8.6 \times 10^{-8} \ ps$,  $\tau_p=2.9 \times 10^{-5} \ ps$ and $\tau_{\epsilon}=7.6 \times 10^{-2} \ ps$ 
}
\label{figmetal}
\end{figure}
\begin{figure}
 \centering
\includegraphics[width=\columnwidth]{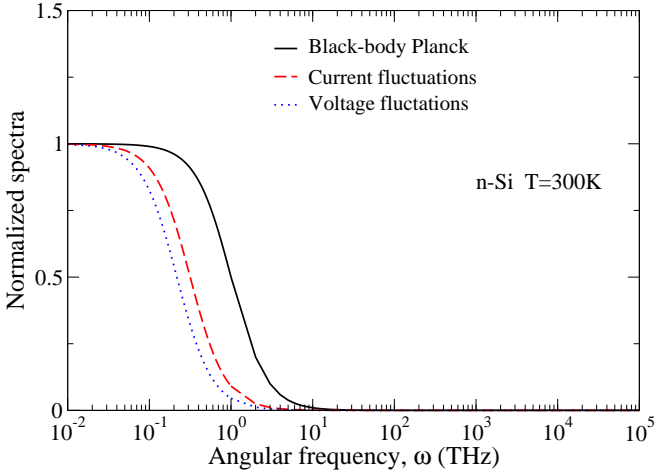}
 \caption{ Noise and black-body spectra near cut-off for the case of a n-type Si sample at $T=300 \ K$ with a donor concentration of $N_D=10^{13} \ cm^{-3}$.
Here typical time scales are, respectively, of $\tau=0.24 \ ps$, $\tau_d=4.6 \ ps$,  $\tau_p=1.1 \ ps$, and $\tau_{\epsilon}=7.6 \times 10^{-2} \ ps$.
}
\label{figsi}
\end{figure}

Figures (\ref{figmetal}) and (\ref{figsi}) report the normalized spectra near the cut-off region associated with: (i) Planck black-body radiation law, (ii) Nyquist spectral-density with the cut-off associated with the intrinsic equivalent-circuitl of the impedance of Fig. (\ref{figec}) for the cases of voltage and current fluctuations, respectively.
Figure (\ref{figmetal}) refers to the case of a typical metal  and  Fig. (\ref{figsi}) to the case of an $n$-Si semiconductor material with a donor concentration of $N_D=10^{13} \ cm^{-3}$ at the same temperature of $300 \ K$.
For the case of a metal, owing to the significant small values of the intrinsic time scales (see the values of scattering,  dielectric relaxation and plasma times reported in the figure caption) the black-body cut-off occurs at a frequency significantly  lower than the values of the cut-off frequencies for the real part of the admittance or impedance spectra. (Notice the spike of the impedance spectrum at the plasma frequency). 
By contrast, for the case of a typical semiconductor, like $n$-type Si, owing to the significant high values of the intrinsic  time scales (see the values reported  in the figure  caption),  the black-body cut-off occurs at a frequency higher than the values of the cut-off frequencies for the real part of its admittance or impedance spectra. (Notice the absence of the spike of the impedance spectrum at the plasma frequency). 
We remark that Figs. (\ref{figmetal}) and (\ref{figsi}) show that the noise spectra of a conductor, being dependent of the small-signal electrical characteristics of the material differ from that of the black-body which by contrast is a universal characteristic.  
In any case, the cut-off frequencies  of the transport coefficients suffice to avoid the ultra-violet divergence of the voltage and current spectra for a given material.
Furthermore, we notice that the Langevin equation for vacuum resistance also implies avoiding ultraviolet catastrophe for the spectral densities of fluctuations because of finite-size effects associated with the transit time between opposite contacts of the electromagnetic fields, as predicted by the energy equipartition laws for the variance of the fluctuating electromagnetic fields.
Figure (\ref{planck_cf}) shows the normalized correlation function associated with the Planck law (continuous curve) which is compared with a normalized Gaussian shape (dashed cure).  
Both curves look significantly broadened with respect to the shape of the delta function, with the Planck correlation-function exhibiting a longer tail than that of the Gaussian shape. 
\par
Figures (\ref{figmetal}) and (\ref{figsi}) report the normalized spectra near the cut-off region associated with: (i) Planck black-body radiation law, (ii) Nyquist spectral-density with the cut-off associated with the intrinsic equivalent-circuitl of the impedance of Fig. (\ref{figec}) for the cases of voltage and current fluctuations, respectively.
Figure (\ref{figmetal}) refers to the case of a typical metal  and  Fig. (\ref{figsi}) to the case of an $n$-Si semiconductor material with a donor concentration of $N_D=10^{13} \ cm^{-3}$ at the same temperature of $300 \ K$.
For the case of a metal, owing to the significant small values of the intrinsic time scales (see the values of scattering,  dielectric relaxation and plasma times reported in the figure caption) the black-body cut-off occurs at a frequency significantly  lower than the values of the cut-off frequencies for the real part of the admittance or impedance spectra. (Notice the spike of the impedance spectrum at the plasma frequency). 
By contrast, for the case of a typical semiconductor, like n-type Si, owing to the significant high values of the intrinsic  time scales (see the values reported  in the figure  caption),  the black-body cut-off occurs at a frequency higher than the values of the cut-off frequencies for the real part of its admittance or impedance spectra. (Notice the absence of the spike of the impedance spectrum at the plasma frequency). 
We remark that Figs. (\ref{figmetal}) and (\ref{figsi}) show that the noise spectra of a conductor, being dependent of the small-signal electrical characteristics of the material differ from that of the black-body which by contrast is a universal characteristic.  
In any case, the cut-off frequencies  of the transport coefficients suffice to avoid the ultra-violet divergence of the voltage and current spectra for a given material.
Furthermore, we notice that the Langevin equation for vacuum resistance also implies avoiding ultraviolet catastrophe for the spectral densities of fluctuations because of finite-size effects associated with the transit time between opposite contacts of the electromagnetic fields, as predicted by the energy equipartition laws for the variance of the fluctuating electromagnetic fields.
\subsection{Fluctuation dissipation theorem for a gas of photons}
For a gas of photons the number of quasi-particles is not conserved in time and thus the chemical potential is zero.
Furthermore the energy dispersion is
\begin{equation}
\epsilon_k= \hbar \omega ,
\end{equation}
and the density of states $g(\epsilon_k)$ is
\begin{equation}
g(\epsilon_k) 
=V_0 \frac{8 \pi}{c^3} f^2
\end{equation}
\par
Accordingly, by definition it is
\begin{equation}
\frac{\overline{N_p}}{V_0}=   \frac{8 \pi \Gamma(3) \zeta(3) }{c^3 h^3}\  (K_BT)^3
=  (2.02 \ 10^7)  \ T^3   \ (m^{-3})
\end{equation}
with $\Gamma(3) \zeta(3) =\int_0^{\infty} x^2/(e^x-1) dx = 2.404$.
\begin{equation}
\frac{\overline{\epsilon_{tot}}} {V_0}=  \frac {8 \pi \Gamma(4) \zeta(4)} {c^3 h^3} \ (K_BT)^4
=  7.57 \ 10^{-16} \ T^4 \ (Jm^{-3})
\end{equation}
with $ \overline{\epsilon_{tot}}$ the internal  average-energy in the considered volume,  $\Gamma(4) \zeta(4) = \pi^4 / 15=6.49$, and
\begin{equation}
\overline{\epsilon_{mode}} =  \frac{\overline{\epsilon_{tot}}} {\overline{N_p}} =
 \ 2.7 \  K_BT
\end{equation}
with $\overline{\epsilon_{mode}}$ the average energy per photon mode, notice that its value is slightly less than that of the classical case  per full relativistic massive-particle of 
$3 \ K_BT$ due to photon Bose-Einstein distribution, and much less than that of the degenerate case per full degenerate Fermions of $3/5 (\mu_0)$.
\par
Since photons are relativistic particles, in vacuum it is
\begin{equation}
\overline{v^{2'}_z} =  c^2 
\end{equation}
\begin{equation}
\overline{\delta N^2_p} = V_0 \frac{8 \pi}{c^3 h^3} \ (K_BT)^3
\int_0^{\infty} \frac{x^2 e^x}{(e^x-1)^2} dx 
= 1.37 \overline{N_p}
\end{equation}
Notice, that the super Poissonian value, with a Fano factor 
$\overline{\delta N_p^2}/\overline{N_p} = 1.37$ due to Bose-Einstein distribution, should be compared with the value $\overline{\delta N^2}/\overline{N} = 1$ of classical massive-particle statistics and the value $\overline{\delta N^2}/\overline{N} = 0$ of full-degenerate massive-particle statistics at $T=0$.
\par
The duality property of the noise sources of a photon gas writes
\begin{equation}
\overline{\delta v_d^2} = c^2 \frac{\overline{\delta N_p^2}}{{\overline{N}_p}^2}
= \frac{1.37 c^2}{\overline{N_p}}  ,
\end{equation}
that implies  the reciprocity relations 
\begin{equation}
\overline{\delta v_d^2} \ {\overline{N_p}}^2 =
\overline{\delta N_p^2}  \ \overline{v^{2'}_x} =    
1.37 c^2 {\overline{N_p}} = cost \ T^3
\end{equation}
\par
We conclude, that for the photon gas the microscopic noise source comes from the fluctuations of the photon number only, and its average value is proportional to the third power of the temperature. 
At zero temperature fluctuations vanish apart from the presence of zero-point contribution which can be considered associated with Bose-Einstein condensation of the photons ground state and which is responsible of the Casimir effect.
However the Casimir effect does not produce fluctuations by itself but a macroscopic quantum-energy associated with the spatial confinement of the physical system.
\par
For the diffusion coefficient and the plasma mobility of the photon gas we find
\begin{equation}
D = \overline{v^{2'}_x} \tau = cL
\end{equation}
\begin{equation}
\mu^2_P = \frac{q^2 \tau_P^2}{m^2 } =
\alpha 4 \pi \epsilon_0 \ \hbar c \frac{L^2}{c^2} \frac{c^4}{(\hbar \omega)^2} =
\frac{4 \pi \epsilon_0 \hbar c}{137} \frac{L^2c^2}{(\hbar \omega)^2}
\end{equation}
with $\alpha = 1/137$ the fine structure constant, that has been used to convert  the  electron charge into the vacuum charge, and $\hbar \omega$ the photon relativistic mass, that replaces the electron mass.
\section{Conclusions}
This review presents a revisitation  of the fluctuation dissipation theorem (FDT)that from an historical point of view is traced back to the discovery of the fundamental laws governing the black-body radiation spectrum. The main objective of the paper is to further stress the unifying microscopic interpretation of the interaction between radiation and matter, and of electrical noise in particular, given by this theorem. 
Main points that received new insights in specific parts  of the paper are briefly summarized in the following list.
\par\noindent
1 - The zero-point energy term, that is present in the quantum formulation of FDT due to Callen and Welton \cite{callen51}, does not contribute to electrical fluctuations. 
By contrast, it is responsible of the Casimir force, a pure quantum-mechanical  macroscopic effect, that by implying a mechanical instability of the physical system, should be exactly balanced by a reaction force to recover stability, a necessary condition  to detect electrical fluctuations.  
The reaction force can be absorbed by the elastic properties of the environment associated with the physical system.
\par\noindent
2 - The role of the different statistical ensembles (microcanonical, canonical and grand-canonical) in the formulation of the FDT has been analyzed.
Accordingly, the microscopic noise sources associated with the  properties of the medium inside the physical system have been individuated.
\par\noindent
3 - We have identified  the intrinsic equivalent-circuit for the impedance (admittance) model of the physical system appropriate to describe the relaxation of voltage (current) fluctuations and thus the intrinsic bandwidth of the classical noise spectral densities described by Nyquist theorem.
\par\noindent
4 - The appropriate Langevin equations for the current or voltage operation modes used to detect noise spectra have been formulated and solved. 
\par\noindent
5 - The duality and reciprocity relations between microscopic noise sources responsible of the so-called thermal agitation of electric charge in conductors
have been investigated.
Here, fluctuations of the total number of carriers inside the physical system are shown to be responsible of current fluctuations detected in the external short circuit, while fluctuations of the carrier drift-velocity are found to be responsible of the voltage fluctuations detected in the open external circuit.
In essence, the duality relations imply  a generalized  Biot-Savart law converting the variance of current fluctuations with the variance of magnetic field fluctuations, see Eq. (\ref{eq14bis}), and a generalized Ohm law converting the variance of drift-velocity fluctuations with the variance of electric field fluctuations, see Eq. (\ref{eq16}).
\par\noindent
6 -  The FDT  has been generalized to the cases of: (i) the ballistic transport regime of charge-carrier dynamics, (ii) the vacuum and, (iii) the quantum case when $K_{B}T$ is substituted by the Planck spectrum, as originally suggested by Nyquist \cite{nyquist28}. 
We noticed, that the noise spectra associated with current and voltage fluctuations of a conductor are intrinsic characteristics of the material under study, and thus differ from the black-body spectrum that by contrast is a universal property. 
\par\noindent
7 - The validity of the FDT has been extended to the case of fractional statistics.
\par\noindent
8 - The FDT has been applied to the case of a photon gas where noise source has been associated with the fluctuations of the instantaneous photon number inside the physical system.

\section*{Acknowledgements}
{Dr. T. Kuhn from M\"unster University, Germany, is thanked for valuable discussions on the subject.}

\end{document}